\documentclass{article}

\PassOptionsToPackage{numbers, compress}{natbib}

\usepackage[preprint]{neurips_2023}

\usepackage[utf8]{inputenc} 
\usepackage[T1]{fontenc}    
\usepackage{hyperref}       
\usepackage{url}            
\usepackage{booktabs}       
\usepackage{amsfonts}       
\usepackage{nicefrac}       
\usepackage{microtype}      
\usepackage[dvipsnames]{xcolor}        
\usepackage{xspace}
\usepackage{subfig, graphicx}
\usepackage{multicol, multirow}
\usepackage{array, wrapfig}
\usepackage[numbers, compress]{natbib}

\usepackage{graphicx}
\usepackage{pythonhighlight}

\newcolumntype{P}[1]{>{\centering\arraybackslash}p{#1}}

\title{Formatting Instructions For NeurIPS 2023}

\author{%
  Bin Xiao$^1$ \quad  Lei Su$^1$ \\
  $^1$Baichuan Inc. \quad \\
   $\{$xiaobin, sulei$\}$@baichuan-inc.com\\ 
}
  
\newcommand{\Sys}{\texttt{ISO}\xspace}

\begin{document}

\title{\Sys: Overlap of Computation and Communication within Seqenence For LLM Inference}

\date{}
\maketitle

\begin{abstract}

In the realm of Large Language Model (LLM) inference, the inherent structure of transformer models coupled with the multi-GPU tensor parallelism strategy leads to a sequential execution of computation and communication. This results in substantial underutilization of computing resources during the communication phase. To mitigate this inefficiency, various techniques have been developed to optimize the use of computational power throughout the communication process. These strategies primarily involve overlapping matrix computations and communications, as well as interleaving micro-batches across different requests. Nonetheless, these approaches either fall short of achieving ideal overlap or impose certain limitations on their application.
To overcome these challenges, this paper introduces a novel strategy for computation-communication overlap that operates at the sequence level. This method not only enhances the degree of overlap but also minimizes the constraints on its applicability. Experimental evaluations conducted using 30b/70b models have demonstrated significant improvements in efficiency. Specifically, the proposed technique has been shown to reduce time consumption by approximately 35\% on 4090 GPU and by roughly 15\% on A800 GPU during the prefill stage of LLM inference.
\end{abstract}

\section{Introduction}
\label{sec:intro}

Generative Large Language Models (LLMs), epitomized by groundbreaking architectures like GPT-2\cite{gpt2}, ChatGPT\cite{chatgpt}, and GPT-3\cite{gpt3}, have revolutionized the landscape of artificial intelligence. These sophisticated models exhibit remarkable versatility, seamlessly transitioning from crafting imaginative narratives to engaging in lifelike dialogues with humans. Their deep comprehension of natural language has elevated human-computer interaction, automating complex tasks that demand nuanced contextual understanding.

As these models have grown in scale and expanded their input sequence lengths, a new challenge has emerged: traditional hardware limitations prevent a single processing unit from handling the immense model weights or kv cache sizes. In response, tensor parallelism\cite{tensor_parallelism} has been introduced, a technique that divides both the model and its kv cache across multiple processing units, enabling them to operate in parallel. Following a phase of parallel computation, a collective communication step is undertaken to consolidate global information. However, this approach introduces a serial dependency between computation and communication phases, leading to substantial underutilization of computational resources during the communication intervals.

\begin{figure}[h]
\centering
\includegraphics[width=0.8\linewidth]{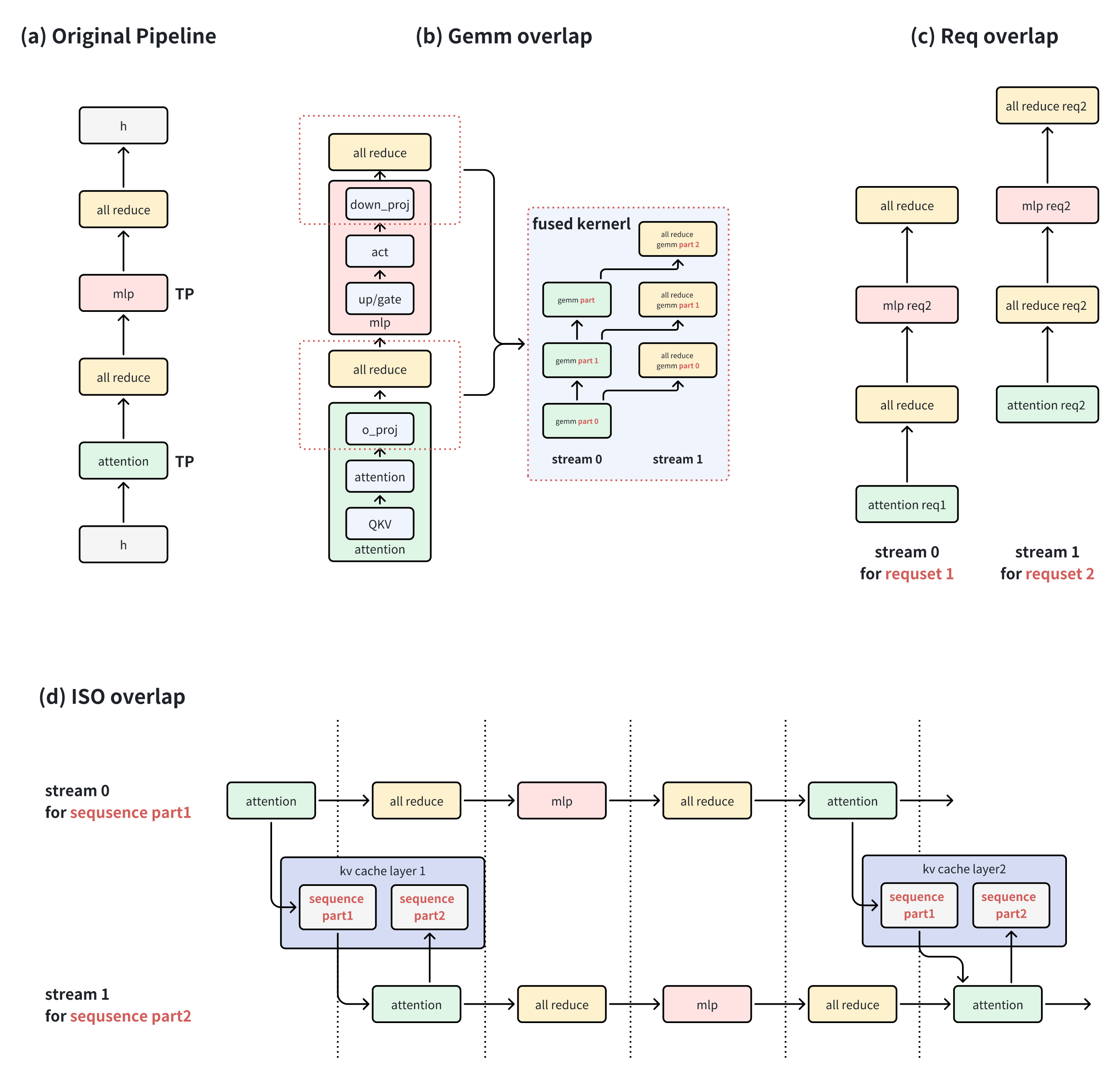}
\caption{Overview of (a) Orginal pipeline, (b) Gemm overlp, (c) Request overlap, (d) ISO overlap.}
\label{fig:intro}
\end{figure}

To address this issue, there are primarily two mainstream solutions, as illustrated in Figure \ref{fig:intro} (b/c). The first solution involves segmenting the matrix computation adjacent to communication into multiple blocks to facilitate overlap \cite{gemm_overlap, gemm_overlap2}. However, in numerous cases, the communication duration exceeds that of the matrix operation, resulting in incomplete overlap. Concurrently overlapping matrix computations with communication can introduce considerable additional computational costs. In specific scenarios, these extra costs may even lead to the strategy yielding negative returns. The second approach entails grouping multiple requests into two micro-batches and executing computation-communication overlap between these micro-batches\cite{req_overlap}. On one hand, it necessitates waiting for at least two requests; on the other hand, the two requests must be as balanced as possible to achieve effective overlap. Simultaneously, while this method can enhance overall throughput, it also results in increased latency for individual requests.

Addressing the previously mentioned challenges, we introduce a strategy that groups micro-batches within a sequence, facilitating overlap between the two, as depicted in Figure \ref{fig:intro} (d). We refer to this method as ISO (Intra-Sequence Overlap). This approach entails dividing the sequence into two parts, prioritizing the computation of the sequence-dependent component (the attention part) before tackling the non-sequence-dependent section. The only requirement is to preserve the order of attention calculations between the two micro-batches.

Similar to other overlap strategies for computation and communication, the decode phase involves minimal computation and communication, rendering overlap unbeneficial. Therefore, our primary focus is on the prefill stage. We evaluated models of approximately 30B and 70B on both the 4090 and A800 platforms. The time consumption was reduced by approximately 35\% on the 4090 and around 15\% on the A800.

In summary, our main contributions include the following points: 
\begin{itemize}
\item We propose a novel computation-communication overlap method for LLM inference; 
\item We tested our strategy on different model sizes and machine types, achieving better results than existing methods;
\item We also discussed the scenarios where this method is applicable and how to optimize it for different scenarios.
\end{itemize}

\section{Background}
\subsection{LLM Inference}

\textbf{Autoregressive decoding}: LLM inference request processing consists of two distinct phases: a prefill phase followed by a decode phase. The prefill phase processes the entire user input prompt and produces the first output token. Subsequently, the decode phase generates output tokens one at a time wherein the token generated in the previous step is passed through the model to generate the next token until a special end-ofsequence token is generated. Note that the decode phase requires access to all the keys and values associated with all the previously processed tokens to perform the attention operation. To avoid repeated recomputation, contemporary LLM inference systems store activations in KV-cache\cite{kv_cache}.

\textbf{Tensor Parallelism (TP)} is a technique used to improve the efficiency of training and running Large Language Models (LLMs) by dividing the model's parameters into smaller segments for parallel processing across multiple devices. This approach reduces memory consumption and enables the use of larger models by distributing computational tasks among several GPUs. A key aspect of tensor parallelism is the integration of inter-card communication mechanisms, which are crucial for maintaining data consistency and coherence across distributed computational resources. During the parallel processing of model components such as attention blocks and multi-layer perceptrons, communication operations are included to ensure consistent results, even when computed across separate devices. Frameworks like Megatron-LM \cite{tensor_parallelism} utilize tensor parallelism, including these communication protocols, to optimize the performance and scalability of LLMs. By efficiently managing data exchange between GPUs, tensor parallelism supports the development and deployment of more powerful language models through improved parallel computation and memory utilization.

\textbf{Chunked-prefills}: The chunked prefill method is engineered to enhance the efficiency of the decoding phase and mitigate delays associated with prolonged prefill computations. As described in SARATHI \cite{chunked_prefill}, this approach divides a prefill request into equally sized computational chunks, enabling the prefill phase of a prompt to be processed across several decoding iterations (each utilizing a subset of the prompt tokens). This strategy facilitates the creation of multiple decode-maximal batches from a single prefill request, thereby optimizing the extent to which decodings can leverage the prefill.

\subsection{Overlap of Computation and Communication}
In large language model (LLM) inference, "Overlap of Computation and Communication" is a crucial optimization technique aimed at enhancing performance and efficiency. This approach addresses the inherent latency between computation and data transmission in distributed systems. 

\textbf{gemm overlap}\cite{gemm_overlap,gemm_overlap2}: The LLM architecture comprises stacked transformer blocks, each integrating attention mechanisms (comprising QKV matrices, attention computations, and an o\_proj matrix) along with MLP components (featuring up/gate matrices, activation functions, and down matrices). Both segments incorporate collective communication processes. The core strategy for optimizing performance involves aligning the execution of specific matrix operations with communication tasks: in the attention component, the o\_proj matrix computation is overlapped with collective communication; similarly, in the MLP segment, the down matrix operation is synchronized with communication activities. This approach aims to enhance efficiency by minimizing idle periods during data exchange across the distributed system.

\textbf{requests overlap}\cite{req_overlap}: is an optimization technique tailored for managing multiple requests in parallel computing settings. It consolidates incoming requests into a unified batch, which is then subdivided into two smaller micro-batches. These micro-batches are intentionally crafted to share overlapping content, enabling one batch to perform computational operations while the other handles communication tasks. By cyclically swapping the functions of these micro-batches between computation and communication, the method ensures efficient use of system resources. This results in diminished idle periods and an uplift in the system's overall throughput. The strategy ingeniously capitalizes on the synergistic relationship between computational and communicational processes, thereby boosting the responsiveness and scalability of distributed systems.
\section{\Sys}

One significant challenge in achieving optimal overlap between matrix computations and communication lies in the fact that, in numerous situations, communication times frequently surpass the duration of adjacent matrix operations, thereby hindering ideal synchronization. At the request level, the prerequisite for multiple concurrent requests may not always be met, particularly in scenarios characterized by low traffic. Furthermore, variations in the computational and communicational demands across different requests can often result in less than perfect overlap. Such overlaps among multiple requests may, in some cases, extend the processing time for individual requests. In light of these challenges, we introduce \Sys as a proposed solution.

\subsection{Overlap strategy}
Figure \ref{fig:intro} depicts our strategy for dividing and overlapping tasks, drawing inspiration from the concept of chunked prefill. This approach involves segmenting a single request into multiple parts along the sequence dimension. Each prefill operation computes only one segment, leveraging the kv cache stored from preceding chunk computations without introducing errors. We split the request into two chunks, facilitating computational and communicational overlap between them. It is essential to maintain the order of computation for chunks within the attention section of the same layer, allowing the second half of the sequence to initiate attention calculations once the first half has completed writing to the kv cache.

\subsection{Asymmetric optimization strategy}
In practical applications, there may be an imbalance in the proportion of computation and communication. In response to this phenomenon, we have implemented some optimization strategies:

\begin{figure}[h]
\centering
\includegraphics[width=0.85\linewidth]{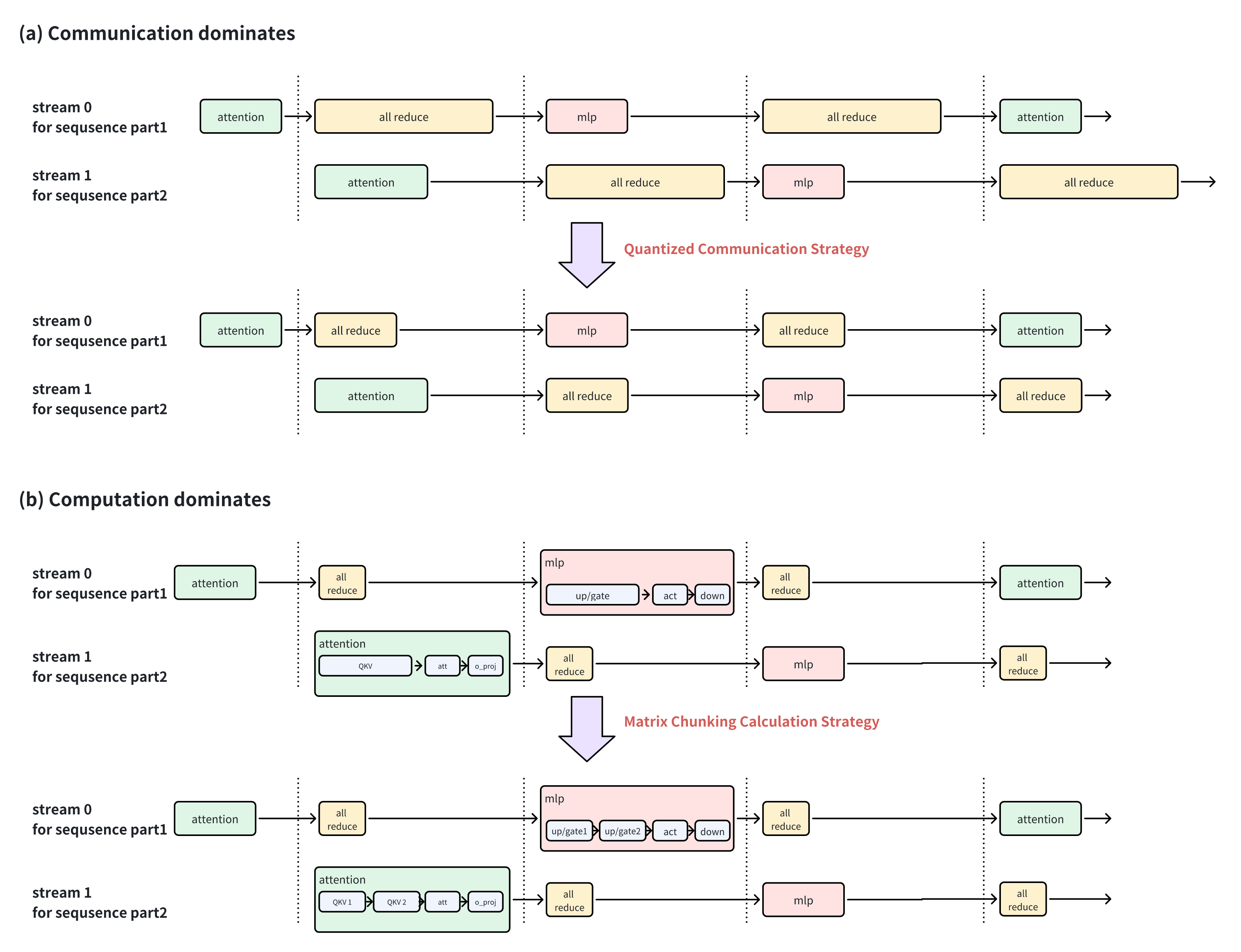}
\caption{Optimization Strategy of (a) Communication dominates, (b) Computation dominates.}
\label{fig:opt}
\end{figure}

\textbf{Communication dominates}: As illustrated in Figure \ref{fig:opt} (a), this scenario predominantly affects the 4090 platform, characterized by its subpar communication capabilities. To mitigate this issue, we have optimized the communication data by converting float16 data to int8 through quantization techniques. This approach has significantly reduced the communication proportion from approximately 75\% to roughly 50\%. Additionally, we are exploring various methods to enhance point-to-point communication efficiency.

\textbf{Computation dominates}: : As shown in Figure \ref{fig:opt} (b), this situation commonly arises in high-performance GPUs, such as the A or H series, where the computation proportion often exceeds 75\%, with longer prompts further increasing this percentage. Concurrently, when accounting for the overlap between computation and communication, NCCL communication tends to consume extra SMs, thereby prolonging the computation time. The impact on A800 ranges from 15\% to 20\%, whereas it is negligible on the 4090 platform. Given the challenge of reducing the computation proportion, our focus has been on minimizing the adverse effects of NCCL communication on computation. Observations reveal that communication seldom surpasses the duration of the initial matrix computation. In response, we have partitioned matrix computations into several segments and employed multiple kernel launches. This strategy ensures that matrix computations can fully exploit the computational power once the communication phase concludes. We are also exploring more effective methods for overlapping gemm computations with communication, as mentioned in Flux\cite{chang2024flux}.
\section{Evaluation}

\subsection{Experiment Settings}

\paragraph{GPU Types, Models, and Baselines} In our experiments, we utilized both the 4090 and A800 GPUs to evaluate dense models with sizes of 30B (MHA) and 70B (GQA). The primary distinction from the baseline configuration lies in the enablement of \Sys, while all other settings remain identical. Our quantization approach entails utilizing int8 for weights, kv cache, and gemm operations, alongside float16 for activations. Notably, the 4090 GPU employs int8 for data transmission.

\paragraph{Metrics} Our primary focus is on comparing the time consumption for the first token processing, specifically for a batch size of 1. Given the sensitivity surrounding internal performance data, we opt not to disclose exact numerical values. Instead, we present the improvements achieved as ratios, providing insights into the relative enhancements without revealing proprietary details.

\subsection{End-to-end Results}

\begin{table}[h]
    \centering
    \small
    \begin{tabular}{|P{1.8cm}|P{0.8cm}|P{0.8cm}P{0.8cm}P{0.8cm}P{0.8cm}P{0.8cm}P{0.8cm}P{0.8cm}P{0.8cm}|}
        \hline
        \multirow{2}{*}{GPU} & \multirow{2}{*}{model}  & \multicolumn{8}{c|}{The percentage decrease in the prefill stage duration at different prompt lengths} \\
        &&1k&2k&4k&8k&16k&32k&64k&128k\\\hline
\rule{0pt}{2.0ex}
        \multirow{2}{*}{4090 4 cards} & \multirow{1}{*}{\centering 30b}  &\textbf{38\%}&\textbf{42\%}&\textbf{43\%}&\textbf{44\%}&\textbf{47\%}&\textbf{48\%}&\textbf{-}&\textbf{-}\\\cline{2-10}
        \rule{0pt}{2.0ex}
        & \multirow{1}{*}{\centering 70b}&\textbf{43\%}&\textbf{44\%}&\textbf{45\%}&\textbf{46\%}&\textbf{47\%}&\textbf{46\%}&\textbf{-}&\textbf{-}\\
        \cline{1-10}
\rule{0pt}{2.0ex}
        \multirow{2}{*}{4090 8 cards} & \multirow{1}{*}{\centering 30b}  &\textbf{11\%}&\textbf{10\%}&\textbf{18\%}&\textbf{21\%}&\textbf{30\%}&\textbf{33\%}&\textbf{36\%}&\textbf{-}\\\cline{2-10}
        \rule{0pt}{2.0ex}
        & \multirow{1}{*}{\centering 70b}&\textbf{14\%}&\textbf{19\%}&\textbf{22\%}&\textbf{23\%}&\textbf{35\%}&\textbf{42\%}&\textbf{39\%}&\textbf{-}\\
        \cline{1-10}
\rule{0pt}{2.0ex}
        \multirow{2}{*}{A800 4 cards} & \multirow{1}{*}{\centering 30b} &\textbf{0\%}&\textbf{8\%}&\textbf{18\%}&\textbf{11\%}&\textbf{12\%}&\textbf{9\%}&\textbf{10\%}&\textbf{5\%}  \\\cline{2-10}
        \rule{0pt}{2.0ex}
        & \multirow{1}{*}{\centering 70b}&\textbf{-6\%}&\textbf{2\%}&\textbf{8\%}&\textbf{10\%}&\textbf{9\%}&\textbf{8\%}&\textbf{8\%}&\textbf{3\%}\\
        \hline

\rule{0pt}{2.0ex}
        \multirow{2}{*}{A800 8 cards} & \multirow{1}{*}{\centering 30b} &\textbf{8\%}&\textbf{24\%}&\textbf{22\%}&\textbf{20\%}&\textbf{16\%}&\textbf{25\%}&\textbf{11\%}&\textbf{10\%}  \\\cline{2-10}
        \rule{0pt}{2.0ex}
        & \multirow{1}{*}{\centering 70b}&\textbf{3\%}&\textbf{9\%}&\textbf{14\%}&\textbf{15\%}&\textbf{16\%}&\textbf{15\%}&\textbf{14\%}&\textbf{7\%}\\
        \hline

    \end{tabular}
    \vspace{10pt}
    \caption{The speedup ratio for the prefill stage duration.}
    \vspace{-10pt}
    \label{tab:e2e}
\end{table}


In real-world operating environments, scenarios involving 1k/2k inputs with a batch size of 1 are uncommon. Generally, batching is utilized, and our main focus is on prompt lengths that exceed 4k. As shown in Table \ref{tab:e2e}, the average improvement observed on the 4090 platform is approximately 35\%, while the A800 platform experiences an improvement of about 15\%. These results stem from our preliminary optimization efforts, and we continue to explore optimizations for less favorable scenarios. The substantial benefit on the 4090 with four cards is attributed to the balanced distribution of computation and communication achieved after adopting int8 for communication. However, when scaling to eight cards on the 4090, the communication ratio escalates, particularly for short prompt lengths. In contrast, the A800 exhibits relatively modest gains due to the dominant role of computation. The concurrent execution of computation and communication can hinder gemm operations, thus limiting the potential improvements. Shorter prompt lengths lead to greater losses from splitting, whereas longer prompts result in a lower communication proportion. We have also experimented with overlapping communication and matrix computations on the A800, yielding marginal gains of 2\%-5\% and even negative gains on the 4090. In all tested scenarios, \Sys surpasses this approach. Regarding request-level overlap, we have yet to implement it, but theoretically, \Sys would offer superior performance by enabling a more uniform distribution of micro-batches across the same requests.


\section{Conclusion}

We introduce a novel method for overlapping computation and communication, which achieves a reduction of approximately 35\% in the duration of the prefill stage on the 4090 platform and about 15\% on the A800 platform, thereby showcasing the efficacy of the \Sys method. Concurrently, we conduct an analysis and comparison of existing approaches for computation-communication overlap, identifying that our methodology outperforms these conventional techniques.

\section{Discussion}
\textbf{The Challenge of Imbalanced Computation and Communication Proportions}

The ideal scenario for maximizing the benefits of overlapping computation and communication is when both components have approximately equal shares. To elaborate further, it is essential to achieve a balance between the attention computation segment and the communication segment, as well as between the MLP computation segment and the communication segment. If there is an imbalance, the overall duration will be determined by the longest segment among these, leading to a decrease in the potential benefits gained from overlapping these processes. 

However, in the case of the A800, computation constitutes a significantly larger portion, and reducing this proportion presents a considerable challenge. It is evident that expanding to 8 cards enhances the effectiveness of ISO. Conversely, the 4090 exhibits a disproportionately large share of communication, which is comparatively easier to mitigate. As previously mentioned, strategies such as quantization or optimizing point-to-point communication at the driver level can be employed. These two scenarios represent rather extreme cases, and newer chips may lie somewhere in between, generally yielding positive gains from ISO.

During our testing, we identified a novel form of imbalance in certain situations. For the A800, both attention and MLP computations far exceed communication, whereas for the 4090, communication surpasses both attention and MLP computations. However, in some specific instances, the time consumed by communication falls between that of attention and MLP. Although this issue is not pronounced in the current context, it is likely to become more significant as the 4090 undergoes optimization. To address this, we propose several potential solutions.

\begin{figure}[h]
\centering
\includegraphics[width=0.95\linewidth]{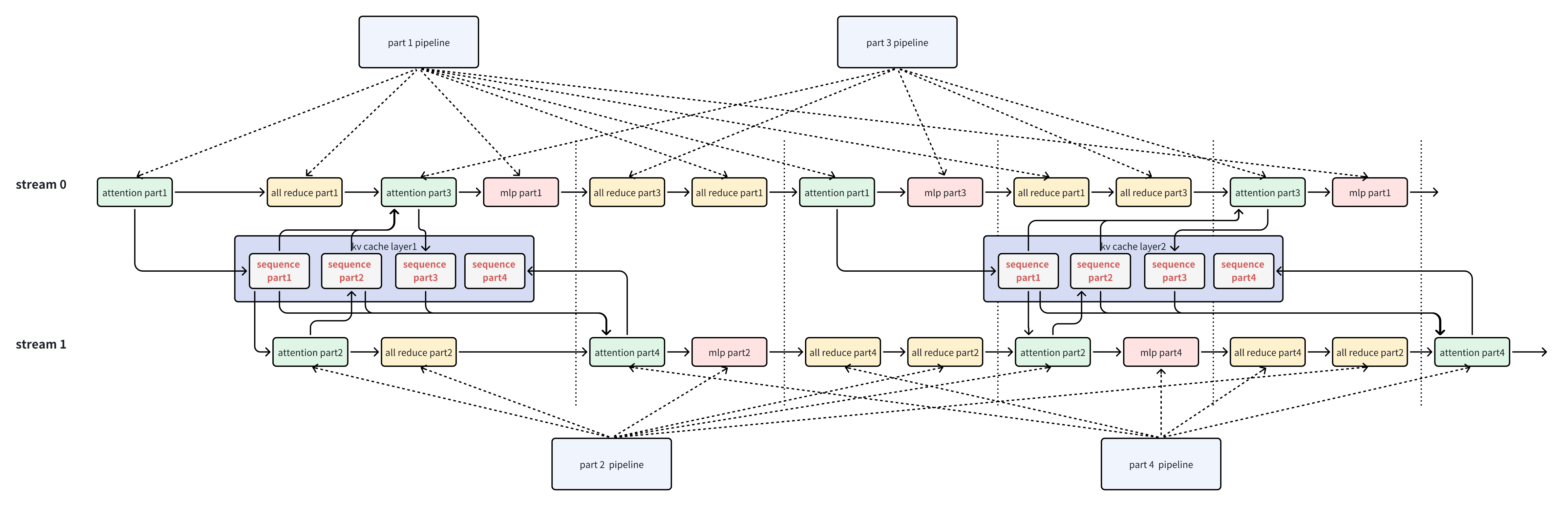}
\caption{Adaptive Attention and MLP Imbalance Splitting Strategy.}
\label{fig:diss_4_part}
\end{figure}

One approach tackles the imbalance within attention itself, where the computational demand of the latter half of the sequence markedly exceeds that of the former half. In such cases, we can adjust the sequence segmentation strategy, for example, by having the first micro-batch compute 60\% of the sequence length and the second compute 40\%. Another solution addresses the imbalance between attention and MLP. Typically, attention computations are more time-consuming, and we can balance the computations of attention and MLP through more intricate micro-batch segmentation strategies, as illustrated in Figure \ref{fig:diss_4_part}.

\textbf{Benefits for the Decode Stage}

Given that the computational and communication loads during the decoding phase are relatively modest, the advantages of overlapping computation and communication are correspondingly limited and may even result in negative returns. Nevertheless, speculative sampling could potentially offer benefits on the 4090 with 4 cards, a direction we are actively investigating. The primary rationale behind this is that speculative sampling involves a greater number of input tokens, thereby increasing the relative computational volume.


\newpage
\bibliographystyle{plain}
\bibliography{egbib}

\begin{thebibliography}{10}

\bibitem{chatgpt}
{ChatGPT: Optimizing Language Models for Dialogue}, 2022.
\newblock \url{https://openai.com/blog/chatgpt/}.

\bibitem{chunked_prefill}
Amey Agrawal, Ashish Panwar, Jayashree Mohan, Nipun Kwatra, Bhargav~S. Gulavani, and Ramachandran Ramjee1.
\newblock {SARATHI: Efficient LLM Inference by Piggybacking Decodes with Chunked Prefills}.
\newblock {\em arXiv preprint arXiv:2308.16369}, 2023.

\bibitem{gpt3}
Tom~B. Brown, Benjamin Mann, Nick Ryder, Melanie Subbiah, Jared Kaplan, Prafulla Dhariwal, Arvind Neelakantan, Pranav Shyam, Girish Sastry, Amanda Askell, Sandhini Agarwal, Ariel Herbert{-}Voss, Gretchen Krueger, Tom Henighan, Rewon Child, Aditya Ramesh, Daniel~M. Ziegler, Jeffrey Wu, Clemens Winter, Christopher Hesse, Mark Chen, Eric Sigler, Mateusz Litwin, Scott Gray, Benjamin Chess, Jack Clark, Christopher Berner, Sam McCandlish, Alec Radford, Ilya Sutskever, and Dario Amodei.
\newblock Language models are few-shot learners.
\newblock In {\em Advances in Neural Information Processing Systems}, pages 1877--1901, 2020.

\bibitem{chang2024flux}
Li-Wen Chang, Wenlei Bao, Qi~Hou, Chengquan Jiang, Ningxin Zheng, Yinmin Zhong, Xuanrun Zhang, Zuquan Song, Ziheng Jiang, Haibin Lin, Xin Jin, and Xin Liu.
\newblock Flux: Fast software-based communication overlap on gpus through kernel fusion, 2024.

\bibitem{req_overlap}
Jiangsu Du, Jinhui Wei, Jiazhi Jiang, Shenggan Cheng, Dan Huang, Zhiguang Chen, and Yutong Lu.
\newblock {Liger: Interleaving Intra- and Inter-Operator Parallelism for Distributed Large Model Inference}.
\newblock {\em https://dl.acm.org/doi/10.1145/3627535.3638466}, 2024.

\bibitem{gemm_overlap}
Abhinav Jangda, Jun Huang, Guodong Liu, Amir Hossein~Nodehi Sabet, Saeed Maleki, Youshan Miao, Madanlal Musuvathi, Todd Mytkowicz, and Olli Saarikivi.
\newblock {Breaking the Computation and Communication Abstraction Barrier in Distributed Machine Learning Workloads}.
\newblock {\em arXiv preprint arXiv:2105.05720}, 2022.

\bibitem{gemm_overlap2}
Suchita Pati, Shaizeen Aga, Mahzabeen Islam, Nuwan Jayasena, and Matthew~D. Sinclair.
\newblock {T3: Transparent Tracking \& Triggering for Fine-grained Overlap of Compute \& Collectives}.
\newblock {\em arXiv preprint arXiv:2401.16677}, 2024.

\bibitem{kv_cache}
Reiner Pope and et~al. Sholto~Douglas.
\newblock {Efficiently scaling transformer inference.}
\newblock {\em Proceedings of Machine Learning and Systems 5 (2023).}, 2023.

\bibitem{gpt2}
Alec Radford, Jeffrey Wu, Rewon Child, David Luan, Dario Amodei, and Ilya Sutskever.
\newblock Language models are unsupervised multitask learners.
\newblock {\em OpenAI blog}, 1(8):9, 2019.

\bibitem{tensor_parallelism}
Mohammad Shoeybi, Mostofa Patwary, Raul Puri, Patrick LeGresley, Jared Casper, and Bryan Catanzaro.
\newblock {Megatron-lm: Training multi-billion parameter language models using gpu model parallelism}.
\newblock {\em arXiv preprint arXiv:1909.08053}, 2019.

\end{thebibliography}


\end{document}